# Adaptive Switching Control of Wind Turbine Generators for Necessary Frequency Response

Yang Liu, *Student Member, IEEE*, Yichen Zhang, *Member, IEEE,* Kai Sun, *Senior Member, IEEE,*
Xiaopeng Zhao, *Senior Member, IEEE*

*Abstract*—**This letter proposes a new control strategy for wind turbine generators to decide the necessity of switches between the normal operation and frequency support modes. The idea is to accurately predict an unsafe frequency response using a differential transformation method right after power imbalance is detected so as to adaptively activate a frequency support mode only when necessary. This control strategy can effectively avoid unnecessary switches with a conventional use of deadband but still ensure adequate frequency response.**

## I. Introduction

Enabling wind turbine generators (WTGs) to provide frequency support other than its normal operating mode is regarded as an effective way to improve frequency responses of power grids with a high wind penetration. Existing literature focused on designing controllers with WTGs for providing frequency support such as inertia emulation, primary frequency response, temporal power injection [1]-[3]. However, a critical issue faced by power system operators is to determine whether WTGs need to be switched to a frequency support mode under a disturbance [4]-[6]. To address this issue, this letter proposes a novel switching control strategy which predicts the safety of a frequency response right after a disturbance by evaluating the derived semi-analytical solutions of system frequency response model over a certain post-disturbance period of interest, and activates frequency support mode only when the frequency response is predicted as unsafe. The case study shows the effectiveness of the proposed strategy.

## II. Proposed Adaptive Switching Control Strategy of Wind Turbine Generators

Existing switching control strategies of WTGs fall into two categories, as shown in (1) and (2), respectively, where $\Delta w(t_0)$ is the frequency drop measured at $t=t_0$, $\Delta w_{db}$ is a preset deadband width on the frequency deviation, $\Delta w_{crt}(\Delta p_d)$ is the critical deadband width, and $\Delta p_d$ is a power imbalance assumed to be known under a disturbance.

$$\begin{aligned}&\text{if } \Delta w(t_0) < \Delta w_{db} : \text{in MPPT mode,}\\&\text{otherwise: in frequency support mode}\end{aligned} \quad (1)$$

$$\begin{aligned}&\text{if } \Delta w(t_0) < \Delta w_{crt}(\Delta p_d) : \text{in MPPT mode,}\\&\text{otherwise: in frequency support mode}\end{aligned} \quad (2)$$

The strategy in (1) is widely adopted by WTGs manufacturers due to its ease of implementation, e.g., the deadband width recommended by GE is $\Delta w_{db}$=0.15HZ for a certain frequency support mode [4],[7]. However, this strategy is not flexible to adapt to different disturbances and the deadband width is often conservatively small such that the unnecessary mode switching of a WTG could be triggered. In comparison, the strategy in (2) can overcome the conservativeness, but it relies on calculating a critical deadband width for each disturbance (e.g., by extensive simulation or by solving an optimization problem [4]), which is a huge computation burden and makes it difficult for real-time implementation.

### A. Proposed Switching Control Strategy

To overcome the conservativeness and the computation burden of existing strategies (1) and (2), this letter proposes a novel switching control strategy in (3), where $\Delta w(t)$, $t \in [t_0,T]$ is the frequency response predictor and $\Delta w_{lim}$ is the safety limit of frequency drop. The detailed derivation of the frequency response predictor is in Section II-C.

$$\begin{aligned}&\textit{if } \Delta w(t) < \Delta w_{\lim} \text{ for } \forall t \in [t_0,T] : \text{in MPPT mode,}\\&\text{otherwise: in frequency support mode}\end{aligned} \quad (3)$$

It is observed from (3) that the proposed strategy uses frequency deviation at $t=t_0$, the time of detecting power imbalance such as loss of generation or load, but further predicts the frequency response over a future time period ($t_0,T$]. Also, the proposed strategy does not need a deadband since it directly compares the predicted frequency response with the safety limit.

### B. System Model

Consider the augmented frequency response model [4],[6] in (4)-(5), where (4) is the classical frequency response model based on center of inertia, and (5) is the reduced order WTG model.

This work was supported in part by the ERC Program of the NSF and DOE under NSF Grant EEC-1041877 and in part by NSF Grant ECCS-1610025 and NSF Grant CMMI-1661615.

Y. Liu and K. Sun are with the Department of Electrical Engineering and Computer Science and X. Zhao is with the Department of Mechanical, Aerospace, and Biomedical Engineering, University of Tennessee, Knoxville, TN 37996 USA (e-mail: yliu161@vols.utk.edu, kaisun@utk.edu, xzhao9@utk.edu). Y. Zhang is with the Energy System Division, Argonne National Laboratory, Lemont, IL 60439 USA (email: yichen.zhang@anl.gov).

$$\Delta\dot{\omega} = \frac{\omega_s}{2H}\left(\Delta p_m - \Delta p_d + \sum_{i=1}^{N}\Delta p_{gen,i} - \frac{D}{\omega_s}\Delta\omega\right)$$

$$\Delta\dot{p}_m = \frac{1}{\tau_{ch}}\Delta p_v - \Delta p_m \quad (4)$$

$$\Delta\dot{p}_v = \frac{1}{\tau_g}\left(-\Delta p_v - \frac{1}{R}\Delta\omega\right)$$

$$\begin{aligned}\Delta\dot{\omega}_{r,i} &= A_i\Delta\omega_{r,i} + B_{1,i}\Delta\dot{\omega} + B_{2,i}\Delta\omega \\ \Delta p_{gen,i} &= C_i\Delta\omega_{r,i} + D_{1,i}\Delta\dot{\omega} + D_{2,i}\Delta\omega\end{aligned} \quad (5)$$

In (4), $\Delta p_m$, $\Delta p_v$ are output power of turbines and governors; $\tau_{ch}$, $\tau_g$ are their time constants, respectively; $D$ and $R$ are damping and droop coefficients; $\omega_s$ is the nominal frequency; $N$ is the number of WTGs and $i$ is the index of WTGs. In (5), $\Delta w_{r,i}$ and $\Delta p_{gen,i}$ are the rotor speed deviation and the increased power generation. Coefficients $A_i$ and $C_i$ are parameters of the WTG model; $B_{1,i}$ and $D_{1,i}$ are parameters for inertia emulation control mode; $B_{2,i}$ and $D_{2,i}$ are parameters for primary frequency control mode. Parameters $B_{1,i}$, $D_{1,i}$, $B_{2,i}$ and $D_{2,i}$ are zero for the MPPT mode and non-zero for frequency support modes. $H$ is a synthetic or equivalent inertia of the WTG [8].

*C. Frequency Response Predictor*

Given system state variables $\mathbf{x}=[\Delta w, \Delta p_m, \Delta p_v, \Delta w_{r,1},\ldots,\Delta w_{r,N}]^T$ and the disturbance at time instant $t=t_0$, this section aims at finding a function $F(\mathbf{x}(t_0))$ in (6) to predict the frequency response over a future time period $(t_0,T]$.

$$\mathbf{x}(t_0 + \Delta t) = F(\mathbf{x}(t_0)), \forall \Delta t \in [0, T - t_0] \quad (6)$$

Since the analytical expression of (6) is generally unavailable, this letter seeks a semi-analytical, approximate solution about the frequency response in the form of power series of time in (7), where $f_k(\mathbf{x}(t_0))$ is the $k^{th}$ order power series coefficients to be solved.

$$\mathbf{x}(t_0 + \Delta t) = \mathbf{x}(t_0) + f_1(\mathbf{x}(t_0))\Delta t + \cdots f_K(\mathbf{x}(t_0))\Delta t^K \quad (7)$$

To effectively solve the power series coefficients $f_k(\mathbf{x}(t_0))$, a differential transformation (DT) method is adopted due to its proved accuracy and efficiency for solving detailed power system dynamic models in our recent work [9]-[11]. It provides various transformation rules such as those in Table I to transform a function $x(t)$ directly to its power series coefficients $x[k]$.

TABLE I
SELECTED TRANSFORMATION RULES OF DT METHOD [8]-[10]

| No. | Original functions | $k^{th}$ order power series coefficients |
|---|---|---|
| 1 | $ax(t) + b$ | $aX[k] + b\delta[k]$, where $\delta[0] = 1; \delta[k] = 0, k \geq 1$ |
| 2 | $x^2(t)$ | $\sum_{m=0}^{k}X[m]X[k-m]$ |
| 3 | $\dot{x}(t)$ | $(k+1)X[k+1]$ |
| ⋮ | ⋮ | ⋮ |

By applying these rules to each term in (4)-(5), a recursive equation about the $k^{th}$ order power series coefficients $f_k(\mathbf{x}(t_0))=[\Delta w[k], \Delta p_m[k], \Delta p_v[k], \Delta w_{r,1}[k],\ldots, \Delta w_{r,N}[k]]^T$ is obtained in (8)-(12).

$$(k+1)\Delta w[k+1] = \frac{\omega_s}{2H}(\Delta p_m[k] - \Delta p_d\delta[k] \\ + \sum_{i=1}^{N}\Delta p_{gen,i}[k] - \frac{D}{\omega_s}\Delta w[k]) \quad (8)$$

$$(k+1)\Delta p_m[k+1] = \frac{1}{\tau_{ch}}\Delta p_v[k] - \Delta p_m[k] \quad (9)$$

$$(k+1)\Delta p_v[k+1] = \frac{1}{\tau_g}\left(-\Delta p_v[k] - \frac{1}{R}\Delta w[k]\right) \quad (10)$$

$$(k+1)\Delta w_{r,i}[k+1] = A_i\Delta w_{r,i}[k] + B_{2,i}\Delta w[k] \\ + B_{1,i}\cdot(k+1)\Delta w[k+1] \quad (11)$$

$$\Delta p_{gen,i}[k] = C_i\Delta w_{r,i}[k] + D_{2,i}\Delta w[k] \\ + D_{1,i}\cdot(k+1)\Delta w[k+1] \quad (12)$$

For details, the derivation of (8) is elaborated below as an example. The remaining equations (9)-(12) are obtained in a similar procedure.

First, the left-hand side of (8) is obtained using the rule 3:
$$\Delta\dot{\omega} \rightarrow (k+1)\Delta w[k+1]$$

Second, the right-hand side of (8) is obtained using rule 1:
$$\Delta p_m - \Delta p_d \rightarrow \Delta p_m[k] - \Delta p_d\delta[k]$$
$$\sum_{i=1}^{N}\Delta p_{gen,i} \rightarrow \sum_{i=1}^{N}\Delta p_{gen,i}[k]$$
$$-\frac{D}{\omega_s}\Delta\omega \rightarrow -\frac{D}{\omega_s}\Delta w[k]$$

$$\frac{\omega_s}{2H}\left(\Delta p_m - \Delta p_d + \sum_{i=1}^{N}\Delta p_{gen,i} - \frac{D}{\omega_s}\Delta\omega\right) \rightarrow$$
$$\frac{\omega_s}{2H}(\Delta p_m[k] - \Delta p_d\delta[k] + \sum_{i=1}^{N}\Delta p_{gen,i}[k] - \frac{D}{\omega_s}\Delta w[k])$$

Finally, (8) is obtained by equating both sides.

With the derived recursive equations (8)-(12), the power series coefficients $f_k(\mathbf{x}(t_0))$ up to any desired order can be derived. Especially, the frequency response at time instant $t=t_0+\Delta t$ is predicted by power series of time in (13).

$$\Delta w(t_0 + \Delta t) = \Delta w[0] + \Delta w[1]\Delta t + \cdots \Delta w[K]\Delta t^K \quad (13)$$

The predicted frequency response (13) is accurate within a certain time period whose length increases with the order $K$. To ensure the accuracy of the predicted frequency response over a longer time period of interest, arbitrary high order $K$ such as 100 to 200 can be easily derived by the DT method. Besides, the multi-time window strategy [9][10] can be used to further enhance the accuracy of (13).

For illustration, the first three terms in the right-hand side of (13) are derived below.

First, initialize $f_0(\mathbf{x}(t_0))=[\Delta w[0], \Delta p_m[0], \Delta p_v[0], \Delta w_{r,1}[0],\ldots, \Delta w_{r,N}[0]]^T = [\Delta w(t_0), \Delta p_m(t_0), \Delta p_v(t_0), \Delta w_{r,1}(t_0),\ldots, \Delta w_{r,N}(t_0)]^T$.

Second, $f_1(\mathbf{x}(t_0))=[\Delta w[1], \Delta p_m[1], \Delta p_v[1], \Delta w_{r,1}[1],\ldots, \Delta w_{r,N}[1]]^T$ is calculated by

$$\Delta w[1] = \frac{\omega_s}{2H}(\Delta p_m(t_0) - \Delta p_d + \sum_{i=1}^{N}\Delta p_{gen,i}[0] - \frac{D}{\omega_s}\Delta w(t_0))$$

$$\Delta p_m[1] = \frac{1}{\tau_{ch}}\Delta p_v(t_0) - \Delta p_m(t_0)$$

$$\Delta p_v[1] = \frac{1}{\tau_g}\left(-\Delta p_v(t_0) - \frac{1}{R}\Delta w(t_0)\right)$$

$$\Delta w_{r,i}[1] = A_i\Delta w_{r,i}(t_0) + B_{2,i}\Delta w(t_0) + B_{1,i}\Delta w[1]$$

$$\Delta p_{gen,i}[0] = C_i\Delta w_{r,i}[0] + D_{2,i}\Delta w[0] + D_{1,i}\cdot \Delta w[1]$$

Third, $f_2(\mathbf{x}(t_0))=[\Delta w[2], \Delta p_m[2], \Delta p_v[2], \Delta w_{r,1}[2],\ldots,\Delta w_{r,N}[2]]^\mathrm{T}$ is calculated similarly.

$$\Delta w[2] = \frac{\omega_s}{2H}(\Delta p_m[1] + \sum_{i=1}^{N}\Delta p_{gen,i}[1] - \frac{D}{\omega_s}\Delta w[1])$$
$$\vdots$$

Finally, $\Delta w(t_0 + \Delta t) = \Delta w[0] + \Delta w[1]\Delta t + \Delta w[2]\Delta t^2$.

### III. CASE STUDY

The proposed strategy is tested on a modified 10-machine 39-bus system [12], where five synchronous generators are replaced by WTGs. The parameters in (4)-(5) are adopted from [4],[6] where $A_i$ = -0.0723, $C_i$ = 0.0127, $H$=4, $\tau_{ch}$ = 0.3, $\tau_g$ = 0.1, $\omega_s$ = 60HZ, $N$ = 5, $D$ = 1, $R$ = 0.05. In the frequency support mode, $B_{1,i}$=-0.6246, $B_{2,i}$=0.1874, $D_{1,i}$=-0.10, $D_{2,i}$=-0.03. Besides, $K$=200, $\Delta w_{db}$=0.2 HZ and $\Delta w_{lim}$=0.5 HZ.

Two scenarios, i.e., a safe scenario with power imbalance of -500 MW and an unsafe scenario with power imbalance of -1000MW, are tested. Fig. 1 and Fig. 2 give the frequency responses in the two scenarios. The proposed strategy activates the frequency support mode only for the unsafe scenario. In contrast, the deadband based strategy would activate frequency support for both the safe and unsafe scenarios. This result shows the proposed strategy can overcome conservativeness compared with the deadband based strategy.

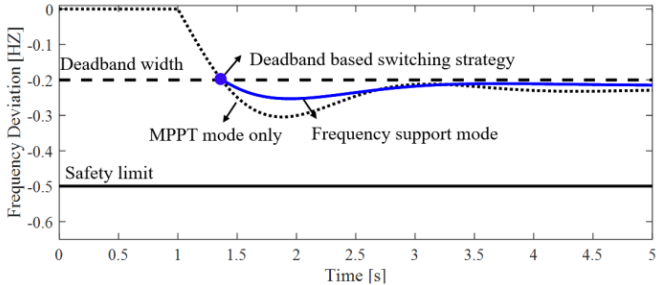

Fig. 1. Switching strategy for the safe frequency response

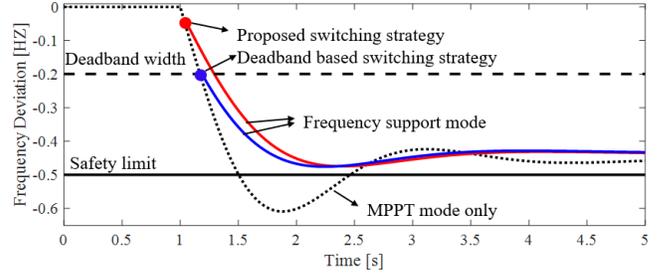

Fig. 2. Switching strategy for the unsafe frequency response

### IV. CONCLUSION

This letter proposes an analytical, adaptive switching control strategy to overcome the conservativeness of the conventional deadband based strategy. For safe responses, the frequency support mode will not be activated even if the deadband is met. For unsafe responses, the frequency support mode is activated immediately once unsafety is predicted. The strategy is shown effective to control WTGs for adequate frequency response.